\documentclass[pdflatex,sn-nature]{sn-jnl}

\usepackage{graphicx}%
\usepackage{multirow}%
\usepackage{amsmath,amssymb,amsfonts}%
\usepackage{amsthm}%
\usepackage{mathrsfs}%
\usepackage[title]{appendix}%
\usepackage{xcolor}%
\usepackage{textcomp}%
\usepackage{manyfoot}%
\usepackage{booktabs}%
\usepackage{algorithm}%
\usepackage{algorithmicx}%
\usepackage{algpseudocode}%
\usepackage{listings}%
\usepackage{subcaption}
\usepackage{float}
\usepackage[final]{changes} 

\theoremstyle{thmstyleone}%
\newcommand{\mb}[1]{\mathbf{#1}}
\newcommand{\pt}{\frac{\partial}{\partial t}}

\theoremstyle{thmstyletwo}%

\theoremstyle{thmstylethree}%

\begin{document}

\title[Article Title]{Reconstructing High-fidelity Plasma Turbulence with Data-driven Tuning of Diffusion in Low Resolution Grids}


\author[1,2]{\fnm{Kunpeng} \sur{Li}}
\equalcont{These authors contributed equally to this work.}
\author[1]{\fnm{Youngwoo} \sur{Cho}}
\equalcont{These authors contributed equally to this work.}
\author[1,3]{\fnm{Xavier} \sur{Garbet}}
\equalcont{These authors contributed equally to this work.}

\author[1]{\fnm{Chenguang} \sur{Wan}}
\author[1]{\fnm{Robin} \sur{Varennes}}
\author[1]{\fnm{Kyungtak} \sur{Lim}}
\author[3]{\fnm{Virginie} \sur{Grandgirard}}

\author*[1]{\fnm{Zhisong} \sur{Qu}} \email{zhisong.qu@ntu.edu.sg}
\author*[2,4]{\fnm{Ong} \sur{Yew Soon}}\email{ASYSOng@ntu.edu.sg}

\affil[1]{\orgdiv{School of Physical and Mathematical Sciences}, \orgname{Nanyang Technological University}, \orgaddress{\city{Singapore}, \postcode{637371}, \country{Singapore}}}

\affil[2]{\orgdiv{College of Computing and Data Science}, \orgname{Nanyang Technological University}, \orgaddress{\city{Singapore}, \postcode{639798}, \country{Singapore}}}

\affil[3]{\orgdiv{CEA}, \orgname{IRFM}, \orgaddress{\postcode{F-13108}, \state{Saint Paul-lez-Durance}, \country{France}}}

\affil[4]{\orgdiv{Centre for Frontier AI Research}, \orgname{Agency for Science, Technology and Research}, \orgaddress{\city{Singapore}, \postcode{138648}, \country{Singapore}}}







\abstract{Developing physically consistent closure models is a longstanding challenge in simulating plasma turbulence, even in minimal systems such as the two-field Hasegawa–Wakatani (HW) model, which captures essential features of drift-wave turbulence with a reduced set of variables. In this work, we leverage theoretical insights from Direct Interaction Approximation (DIA) to construct a six-term closure structure that captures the dominant turbulent transport processes—including both diffusion and hyper-diffusion. While the mathematical form of the closure is fully prescribed by DIA, the corresponding transport coefficients are learned from data using physics-informed neural networks (PINNs). The resulting Extended HW model with Closure (EHW-C) model reveals several nontrivial features of plasma turbulence: notably, some inferred coefficients become negative in certain regimes, indicating inverse transport—a phenomenon absent in conventional closure models. Moreover, EHW-C model accurately reproduces the spectral and flux characteristics of high-resolution Direct Numerical Simulations (DNS), while requiring only one-eighth the spatial resolution per direction—yielding a tenfold speed-up. This work demonstrates how theory-guided machine learning can both enhance computational efficiency and uncover emergent transport mechanisms in strongly nonlinear plasma systems.}




\maketitle

\section{Introduction}\label{sec1}
To achieve burning plasma in tokamaks, it is crucial to improve confinement by estimating and \replaced{controlling}{suppressing} the turbulent transport.
It is \replaced{difficult} {almost impossible}to directly measure such turbulent transport during experiments. 
Therefore, analyzing turbulent transport levels in experiments requires performing direct numerical simulations (DNS) of plasma turbulence \replaced{and validating them against available measurements}{based on measured data}.

However, the computational demands of DNS are formidable because of numerical complexity and requirement of the fine spatiotemporal resolutions.
As an alternative, large-eddy simulation (LES) offers substantially greater efficiency by operating on coarser grids, yet theory-driven closure models remain underexplored in plasma turbulence.
This gap stems partly from the inherent challenges in formulating theory-based closure models and the difficulty in predicting the proper closure in given condition.
Even within the reduced-complexity turbulent system, like Hasegawa–Wakatani (HW) equations\cite{bib3}, conventional closures (e.g., the Smagorinsky model\cite{Smagorinsky1963}) \replaced{are not enough}{fail} in recovering key metrics such as particle flux and turbulence spectra \cite{bib4}. 

Recent advances in machine learning, demonstrated across a wide range of applications\cite{farcacs2025distributed,farcas2023parametric,qian2022reduced,issan2023predicting,clavier2025generative}, have led to the development of data-driven closure models to address these challenging problems. 
For instance, a closure framework utilizing StyleGAN \cite{karras2019style} helps reduce the computational cost of DNS simulations, but it is primarily limited to generating improved initial conditions by replacing conventional random perturbations with more physically meaningful states\cite{bib4}. 
In the case of purely data-driven subgrid closure models, physical constraints on energy and flux have been incorporated into the loss function to achieve partial improvement in conservation\cite{artigues2025accelerating}. 
However, this black-box closure model still faces difficulties in ensuring strict conservation of mass, momentum, and energy, and lack intuitive interpretability and physical transparency.
Additionally, while attempts have been made to compare various alternative schemes, none have simultaneously reproduced the complete suite of energy spectra, particle flux, and enstrophy dynamics\cite{bib5}. 
These limitations highlight the pressing need for closure models that are both computationally tractable and physically interpretable.

To bridge this gap, we introduce the Extended HW with Closure (EHW-C) model—a fully physics-based closure model that combines analytic derivation with AI-driven inference and can be integrated directly into existing simulators. 
When implemented, EHW-C delivers quantitative agreement between coarse-grid and fine-grid simulations, reproducing high-resolution fidelity without relying on any precomputed data. 
Developing a closure model that is both comprehensive and physically self-consistent remains a significant challenge. 
To overcome this, we employ a hybrid strategy: the functional forms of the closure terms are grounded in insights from Direct Interaction Approximation (DIA) theory\cite{Kraichnan1958}, while the free coefficients of these terms are determined through a Physics-Informed Neural Networks (PINNs) framework\cite{bib6}.

Under reasonable assumptions, DIA predicts a closure structure consisting of six coupled diffusion and hyper-diffusion terms linking the density and vorticity fields. 
These include both self-diffusion and cross-diffusion effects, the latter capturing the influence of density fluctuations on vorticity evolution and vice versa. 
While often neglected in simplified models, such cross-terms are predicted by DIA to play a significant role in the nonlinear energy transfer processes characteristic of plasma turbulence.

\begin{figure}[h]
\centering
\includegraphics[width=1\textwidth]{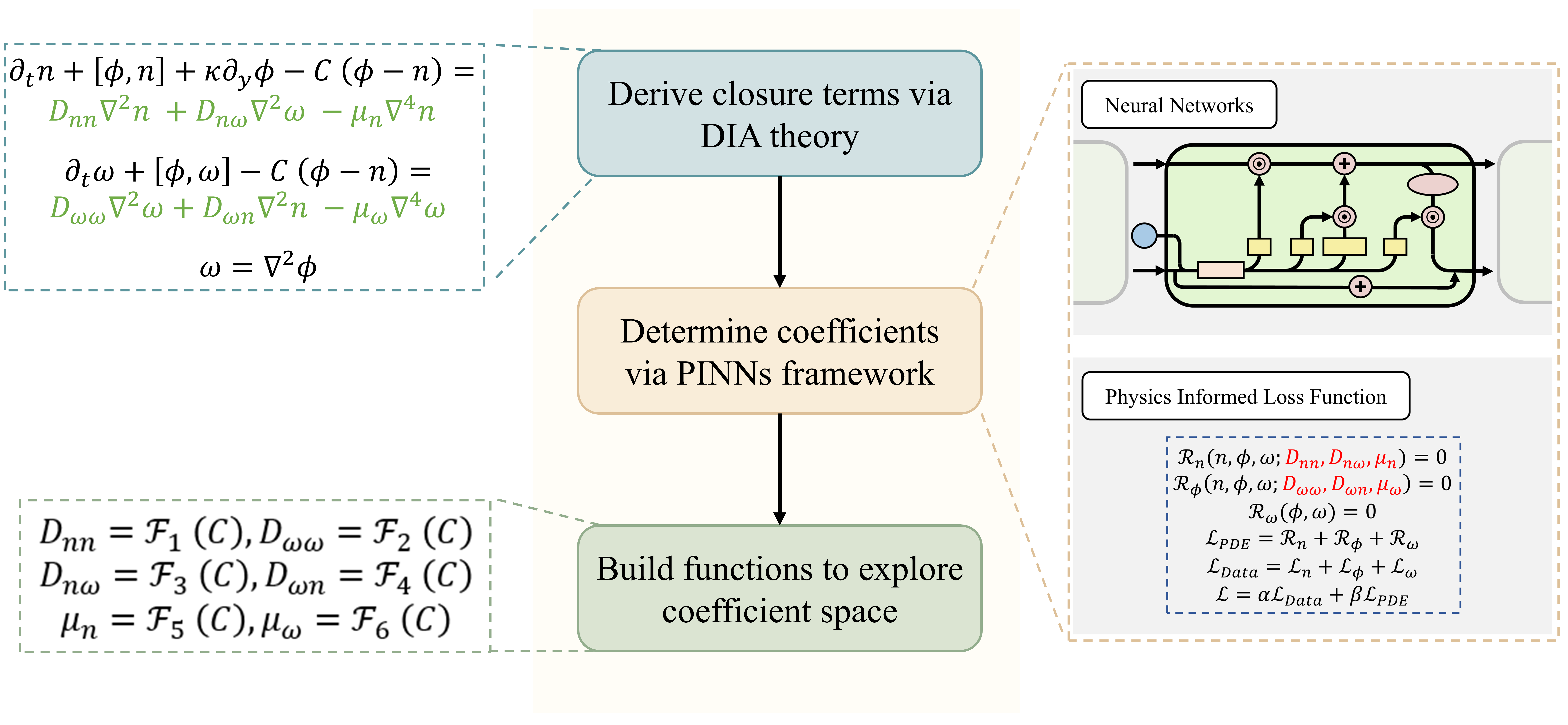}
\caption{Overall workflow of this study. Closure terms are first derived from the DIA theory, yielding a extended form of the Hasegawa–Wakatani equations with unknown dissipation and coupling coefficients. These coefficients are then identified using a PINNs framework, where a neural network is trained under hybrid real-space and spectral-space constraints to enforce consistency with the governing equations. Once identified, the coefficients are expressed as functions of the adiabatic parameter $C$, enabling the construction of functions that capture the dependence of closure behavior on system parameters.}\label{fig:workflow}
\end{figure}

PINNs have emerged as a powerful paradigm in scientific machine learning and have become a promising method for solving complex problem involving partial differential equations (PDEs)\cite{karniadakis2021physics,winovich2019convpde,YeonjongCommunicationsinComputationalPhysics,PatelJCP}. 
By embedding the governing equations directly into the loss function alongside boundary and initial conditions, PINNs seamlessly integrate fundamental physical laws into the network architecture, thereby recasting traditional PDE formulations as equivalent constrained optimization tasks. 
For inverse problems, the unknown parameters within the PDEs can be treated as additional trainable variables.
In this way, PINNs can \replaced{allow a}{easily} recovery of latent physics and solutions in a single end-to-end learning loop. 
This capability makes PINNs particularly well suited to the context of closure‐model coefficient identification in turbulence modeling involved in this work. 
Classical closure schemes introduce closure terms whose form may be motivated by dimensional analysis, asymptotic expansions or other theoretical derivation but whose coefficients must be calibrated against high‐fidelity data (for example, Smith and Hammett's work for the single field Hasegawa-Mima equation\cite{smith1997eddy}). 
In contrast, by embedding the unknown closure coefficients as trainable parameters and minimizing the PDE‐residual loss against available observations, PINNs perform a data‐driven optimization that enforces physical laws exactly while automatically recovering the coefficients that can reproduce the observed flow statistics. 
As training proceeds, gradient‐based optimization updates both the network weights and the closure coefficients until the composite loss converges, yielding a self‐consistent closure model. 
A growing body of work illustrates the versatility and robustness of PINNs for such inverse tasks\cite{rossi2023potential,parfenyev2024inferring,berardi2025inverse}. 
Inspired by PINNs’ proven effectiveness in PDE‐based inverse problems, we employ the physics‐informed neural networks to identify our closure‐model coefficients in an end-to-end, data-driven framework.

In summary, to establish \added{the} EHW-C model, we first derive the specific function form of the closure terms based on DIA theory to retain full physical interpretability, and then leverage PINNs to recover optimal coefficients directly from observation data. The overall workflow of the study is illustrated in Fig.~\ref{fig:workflow}. Crucially, our results reveal that some of the inferred closure coefficients exhibit negative values in specific parameter regimes, suggesting the presence of inverse transport—a mechanism absent in conventional diffusive models. When validated on coarse-grid DNS, EHW-C model achieves excellent agreement with high-resolution simulations while reducing computational cost by over 90\%. Moreover, by expressing the inferred coefficients as smooth functions of the adiabatic parameter, we generalize the model beyond the training set, enabling rapid and accurate simulations across a wide range of turbulence regimes. This work demonstrates how theory-guided machine learning can unlock physically consistent closure models that are both efficient and capable of capturing emergent phenomena in plasma turbulence.

\section{Results and Discussion}\label{sec2}
\subsection{Development of Closure Terms}\label{subsec2}
The database for training the PINN model is constructed based on the Hasegawa-Wakatani equations\cite{bib3}. 
The exact form of the equation we used is as follows:
\begin{subequations}
\begin{align}
    \pt n + [\phi,n] + \kappa \,\frac{\partial}{\partial y} \phi &= C \bigl(\phi - n\bigr) + D_{0}\,\nabla^2 n.  \label{hw_1}\\
    \pt \omega + [\phi,\omega] &= C \bigl(\phi -  n\bigr) + D_{0}\,\nabla^2 w . \label{hw_2}
\end{align}
\end{subequations}
Here, $n$ is the density fluctuation, $\phi$ is the electrostatic potential, $\omega=\nabla^{2}\phi$ is the vorticity, $\kappa=-\partial_{x} N$ is the gradient of equilibrium density $N=N(x)$ in $x$ direction, and $C$ is the adiabaticity parameter.
The system is in the hydrodynamic limit when $C\ll 1$, whereas it is in the adiabatic limit when $C\gg1$.
$[A,B]=\partial_{x}A\partial_{y}B-\partial_{x}B\partial_{y}A$ is the Poisson bracket.
The second terms of R.H.S of Eqs. (\ref{hw_1}) and (\ref{hw_2}) are the dissipation terms.

Low-resolution simulations lose information about short wavelength fluctuations.
A key piece of this lost information is the nonlinear interaction between long and short wavelength fluctuations.
Direct Interaction Approximation (DIA) method provides a set of evolution equations for cross-correlation functions at wavenumber $k$. 
\replaced{These equations are still difficult to solve and are simplified by assuming a Markovian process, and thus exponentially decaying correlation functions This method is called eddy-damped quasinormal Markovian (EDQNM) approximation\cite{Orszag1970}.}{A widely adopted simplification of DIA is the eddy-damped quasinormal Markovian (EDQNM) approximation\cite{Orszag1970}, which assumes a Markovian process leading to exponentially decaying correlation functions.}

This method has also been applied in plasma physics based on Hasegawa-Wakatani equations, to study the nonlocal behavior of turbulence.
In this work, we introduce the Extended Hasegawa-Wakatani with Closure (EHW-C) model, shown as Eqs. (\ref{hw_c1}) and (\ref{hw_c2}), which directly incorporates closure terms to describe the nonlinear interaction involving short wavelength fluctuations.

\begin{subequations}
\begin{align}
    \pt n + [\phi,n] + \kappa \,\frac{\partial}{\partial y} \phi &= C \bigl(\phi - n\bigr) + D_{nn}\,\nabla^2 n + D_{n\omega}\,\nabla^2 \omega - \mu_n\nabla^4 n. \label{hw_c1}\\
    \pt \omega + [\phi,\omega] &= C \bigl(\phi -  n\bigr) + D_{\omega\omega}\,\nabla^2 w + D_{\omega n}\,\nabla^2 n - \mu_\omega\nabla^4 \omega. \label{hw_c2}
\end{align}
\end{subequations}
Compared to Eqs. (\ref{hw_1}) and (\ref{hw_2}), the second to fourth terms on the R.H.S. of Eqs. (\ref{hw_c1}) and (\ref{hw_c2}) are the closure terms derived from DIA methods.
Specifically, $D_{nn}$ and $D_{\omega\omega}$ are the diffusion coefficients for density and vorticity fluctuations, respectively.
$D_{n\omega}$ and $D_{\omega n}$ are the cross terms, representing stresses between the density by the vorticity fields.
$\mu_{n}$ and $\mu_{\omega}$ are the hyper-diffusion terms.
Although analytical expressions for these coefficients have been derived in previous works\cite{Gang1990,Gurcan2006}, they require exact information about the nonlinear interaction from short wavelength fluctuations.
Therefore, directly measuring these coefficients from LES and applying them to low resolution simulations in challenging.
In this work, we utilize PINNs to determine these coefficients.

\added{It should be noted that in this study, we employ the standard Hasegawa-Wakatani equations rather than modified versions that include zonal flow dynamics. 
This choice is motivated by the inherent difficulty in incorporating zonal flow effects within the DIA closure derivation framework. 
The complex interactions between zonal flows and turbulent fluctuations pose theoretical challenges for deriving consistent closure terms from first principles using DIA theory. 
Furthermore, by excluding zonal flow dynamics, we aim to achieve a more precise understanding of closure models that govern enstrophy cascade processes in the absence of zonal flow structures. 
This framework allows us to focus on the fundamental turbulent transport mechanisms and validate our physics-informed approach under well-controlled conditions before extending to more complex systems that include zonal flow dynamics.}

\added{For this purpose,} TOKAM2D\cite{tokam2d2018, tokam2d2022} code is employed to generate the training data by Hasegawa-Wakatani equations and to perform simulations of EHW-C model.
The high resolution simulations use $512\times512$ grids, corresponding to \replaced{cutoff}{maximum} wavenumber $k_{\added{cutoff}}\sim15$. 
In contrast, the low resolution simulations and the simulations for EHW-C model utilize $64\times64$ grids, for which the \replaced{cutoff}{maximum} wavenumber $k_{\added{cutoff}}\sim 2.5$.
To investigate the changes in nonlinear interaction more clearly, the model is trained for a range of adiabatic parameters $C$.

\subsection{Model Selection}\label{subsec2}

To identify the most suitable neural architecture for the PINNs framework, we systematically evaluate a set of candidate models known for their effectiveness in capturing spatiotemporal dynamics. 
The candidates include three time series data prediction models: convolutional long short-term memory networks (ConvLSTM) \cite{shi2015convolutional}, predictive recurrent neural networks (PredRNN) \cite{wang2022predrnn}, and Trajectory GRU (TrajGRU) \cite{shi2017deep}. 
ConvLSTM integrates convolutional operations within the gating mechanisms of LSTM units, enabling efficient modeling of both spatial and temporal correlations. 
PredRNN introduces memory-in-memory mechanisms to better retain long-term dependencies and improve sequence prediction accuracy. 
TrajGRU employs learnable, location-variant recurrent connections to explicitly model non-stationary spatiotemporal dynamics. 
Based on comparative assessment of their consistency with physical constraints and computational efficiency, we select the optimal model configuration for integration into the PINNs framework.

To conduct this comparative evaluation, we employ an adiabatic coefficient of $C$=1.0 as a representative test case and train all candidate models using an identical dataset.
Specifically, the closure coefficients inferred by each candidate model—namely, $\{ D_{nn}, D_{n\omega}, \mu_n, D_{\omega\omega}, D_{\omega n}, \mu_\omega \}$—are incorporated into the TOKAM2D code to perform direct numerical simulations.  
The resulting physical observables, including density and potential spectra, are then used as benchmarks to assess the accuracy and effectiveness of each model’s inferred coefficients. 
The procedures for data generation, as well as the specific implementation details of the PINNs framework, are described in the \deleted{Methods} section \added{\ref{subsec2}}.

\begin{figure}[h]
\centering
\includegraphics[width=1\textwidth]{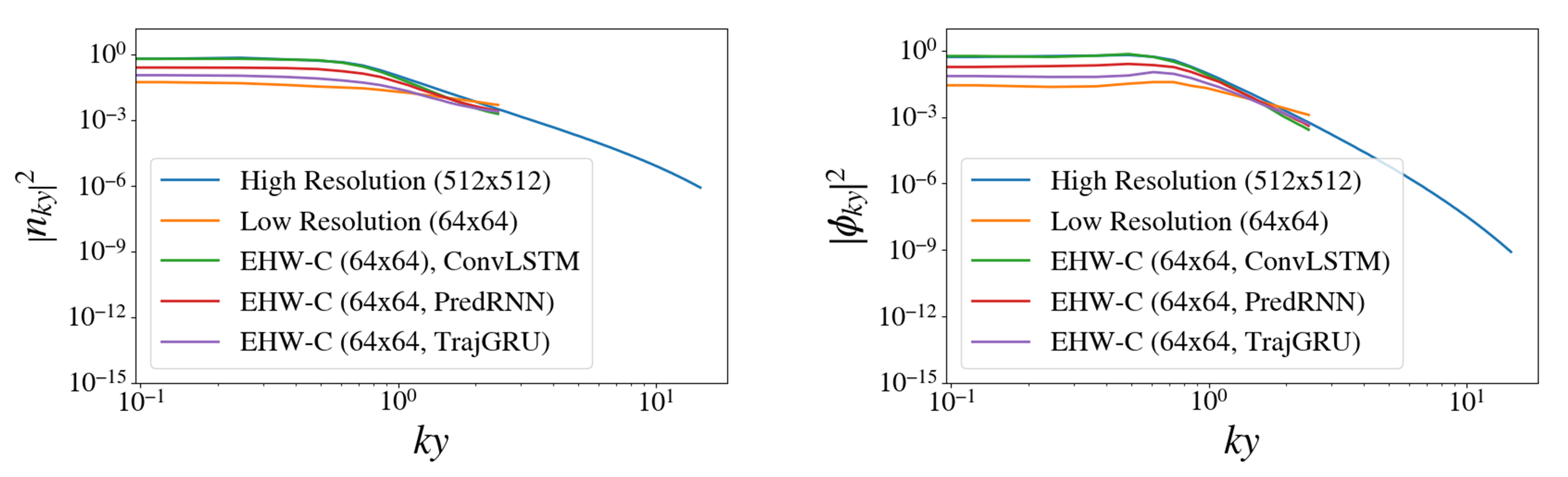}
\caption{Comparative model performance in reproducing turbulence spectra. Spectral comparisons of the density $|n_{ky}|^2$ (left) and potential $|\phi_{ky}|^2$ (right) between high-resolution DNS (512×512), low-resolution DNS (64×64), and EHW-C models at same low-resolution, where the coefficients are determined by different spatiotemporal neural architectures. Among the tested architectures—ConvLSTM, PredRNN and TrajGRU—ConvLSTM yields spectra in closest agreement with the DNS benchmark, demonstrating superior ability to recover physically consistent coefficients.}\label{fig:compare}
\end{figure}

As shown in Fig.~\ref{fig:compare}, the high-resolution simulation ($512\times512$) exhibits characteristic power law decay in both density and potential spectra at high wavenumbers.
\added{Here, $|n_{k_y}|^{2} = \Sigma_{k_x} |n_{k}|^{2}$ and $|\phi_{k_y}|^{2} = \Sigma_{k_x} |\phi_{k}|^{2}$.}
In contrast, the low-resolution simulation ($64\times64$) without closure severely underestimates the spectral power at all wavenumbers, particularly failing to capture the proper scaling behavior.
All three EHW-C models with different neural architectures-ConvLSTM, PredRNN, and TrajGRU-demonstrate significant improvement over the \deleted{unclosed} low-resolution case.
Notably, ConvLSTM shows the closest agreement with the high-resolution reference, accurately reproducing both spectral amplitude and the power law scaling across the entire wavenumber range.
While PredRNN and TrajGRU also enhance the low-resolution predictions, they exhibit slightly larger deviations from the high-resolution benchmark, particularly in the low and intermediate wavenumber regimes.
Based on these quantitative comparisons, we select ConvLSTM as the backbone network for the PINNs framework in subsequent sections.

\subsection{Verification and Validation of Identified Coefficients}\label{subsec2}

We first evaluate the performance of the proposed EHW-C model, in which the closure coefficients are identified using the PINNs framework. 
For each selected adiabatic coefficient value $C$, six closure coefficients are inferred, as summarized in Table~\ref{coefficients}. 
In the near-hydrodynamic limit ($C=0.2$), $D_{nn}$ is positive while $D_{n\omega}$ and $\mu_n$ are negative. 
This reflects the physical characteristics of this limit, where energy transfer to \added{the electric potential} $\phi$ is weak, so the equilibrium spectrum of density is determined by the interaction among the linear drive, diffusion, and hyperdiffusion.
Meanwhile, $D_{\omega\omega}$ is negative but $\mu_{\omega}$ is positive regardless of $C$.
This indicates the presence of negative vorticity diffusion, which introduces an inverse cascade, and strong damping of high-$k$ by hyperdiffusion.
This behavior is consistent with the characteristics of 2D fluids\cite{Kraichnan1976}.
Both $D_{n\omega}$ and $\mu_{n}$ become positive for higher $C$, which means that the coupling between density fluctuations and $\omega$ becomes significant in this regime. 
When the system approaches the near-adiabatic limit, i.e. $C=5$, EHW-C model can be reduced to Hasegawa-Mima equation \cite{hasegawa1977stationary,hasegawa1978pseudo} with our closure terms.
\begin{equation}
    \pt (\nabla^{2}\phi - \phi)  + [\phi,\nabla^{2}\phi] -\kappa\frac{\partial}{\partial y}\phi = -(D_{nn}-D_{\omega n})\nabla^{2}\phi -(D_{n\omega}-D_{\omega\omega}-\mu_{n})\nabla^{4}\phi - \mu_{\omega}\nabla^{6}\phi \label{HM1}
\end{equation}
Here, the first term in the R.H.S. of Eq. (\ref{HM1}) is the frictional damping term, since $D_{nn}-D_{\omega n}=1.04\times10^{-2}>0$.
The second term is the negative diffusion, which has a positive coefficient ($D_{n\omega}-D_{\omega\omega}-\mu_{n}=1.19\times10^{-3}>0$).
The last term is the hyperdiffusion term, with a coefficient of $\mu_{\omega}=8.79\times10^{-3}>0$.
This estimation shows good agreement with the work of Smith and Hammett\cite{smith1997eddy} on Hasegawa-Mima equation, which exhibits weak negative viscosity in the low-$k$ regime, and \replaced{damping}{positive viscosity} at high-$k$ regime due to hyperviscosity.
Therefore, we can conclude that our PINNs framework has effectively learned the coefficients required to produce a stable and physically valid spectrum in this particular limit.

\begin{table}[h]
\caption{Identified closure coefficients for different adiabaticity parameter \( C \).}
\label{coefficients}%
\begin{tabular}{@{}llllll@{}}
\toprule
$C$ & \(0.2\) & \(1.0\) & \(2.0\) & \(3.0\) & \(5.0\) \\
\midrule
$D_{nn}$   & $6.06 \times 10^{-2}$ & $4.27 \times 10^{-2}$ & $2.36 \times 10^{-2}$ & $1.58 \times 10^{-2}$ & $5.96 \times 10^{-3}$ \\
$D_{n\omega}$  & $-4.00 \times 10^{-3}$ & $-5.70 \times 10^{-4}$ & $1.90 \times 10^{-4}$ & $9.00 \times 10^{-4}$ & $1.24 \times 10^{-3}$ \\
$\mu_n$    & $-3.40 \times 10^{-3}$ & $4.10 \times 10^{-4}$ & $2.89 \times 10^{-3}$ & $3.92 \times 10^{-3}$ & $4.46 \times 10^{-3}$ \\
$D_{\omega\omega}$   & $-4.30 \times 10^{-2}$ & $-5.11 \times 10^{-2}$ & $-5.04 \times 10^{-2}$ & $-3.06 \times 10^{-2}$ & $-4.41 \times 10^{-3}$ \\
$D_{\omega n}$   & $-1.37 \times 10^{-2}$ & $-8.51 \times 10^{-3}$ & $-7.04 \times 10^{-3}$ & $-5.04 \times 10^{-3}$ & $-4.39 \times 10^{-3}$ \\
$\mu_\omega$   & $2.10 \times 10^{-2}$ & $2.19 \times 10^{-2}$ & $1.60 \times 10^{-2}$ & $1.30 \times 10^{-2}$ & $8.79 \times 10^{-3}$ \\
\botrule
\end{tabular}
\end{table}

To comprehensively evaluate the generalization capability of the proposed EHW-C model, we conduct comparison experiments using the identified closure coefficients under various adiabatic coefficients \( C = \{0.2,1.0,2.0,3.0,5.0\} \). 
The representative results are summarized in Fig.~\ref{fig:flux_subfigs}, with additional simulation results provided in the supplementary material. 
Comparisons are made among high-resolution DNS results ($512 \times 512$), low-resolution DNS simulations without closure ($64 \times 64$), and simulations using the proposed EHW-C model at the same low-resolution ($64 \times 64$).

For the near-hydrodynamic regime ($C=0.2$), the flux evolution shows that the low-resolution simulation without closure severely underestimates the particle flux, remaining nearly constant at a low value throughout the simulation time. 
In contrast, EHW-C model successfully recovers the fluctuating behavior observed in the high-resolution reference, capturing both the magnitude and temporal variations of the particle flux. 
The spectral analysis reveals that while the \deleted{unclosed} low resolution simulation fails to reproduce the proper energy cascade, particularly in the density spectrum $|n_{k_{y}}|^{2}$, EHW-C model accurately reconstructs the spectral slopes across all three quantities ($|n_{k_{y}}|^{2}$, $|\phi_{k_{y}}|^{2}$, and $|n_{\omega}|^{2}$). 
Most notably, EHW-C model correctly captures the steeper decay in the potential spectrum, which is a signature characteristic of this regime. 
\added{To validate the physical consistency of the observed spectral behavior, we analyze the energy spectrum of HW system, defined as $E_{k}=\frac{1}{2}(|n_{k}|^{2}+k^{2}|\phi_{k}|^{2})$.
In this case, the density spectrum exhibits $|n_{k}|^{2}\propto k^{-2.48}$, indicating that density fluctuations dominates over electrostatic potential fluctuations in the energy spectrum.
Consequently, the overall energy spectrum follows $E_{k}\sim\frac{1}{2}|n_{k}|^{2}\propto k^{-2.48}$, which aligns well with theoretical expectations for direct enstrophy cascade in 2D fluid turbulence\cite{bib2,bib3}.
This spectral consistency reinforces the physical validity of EHW-C model in capturing the correct energy transfer mechanisms of the near-hydrodynamic regime.
}

\begin{figure}[H]
\centering
\includegraphics[width=1\textwidth]{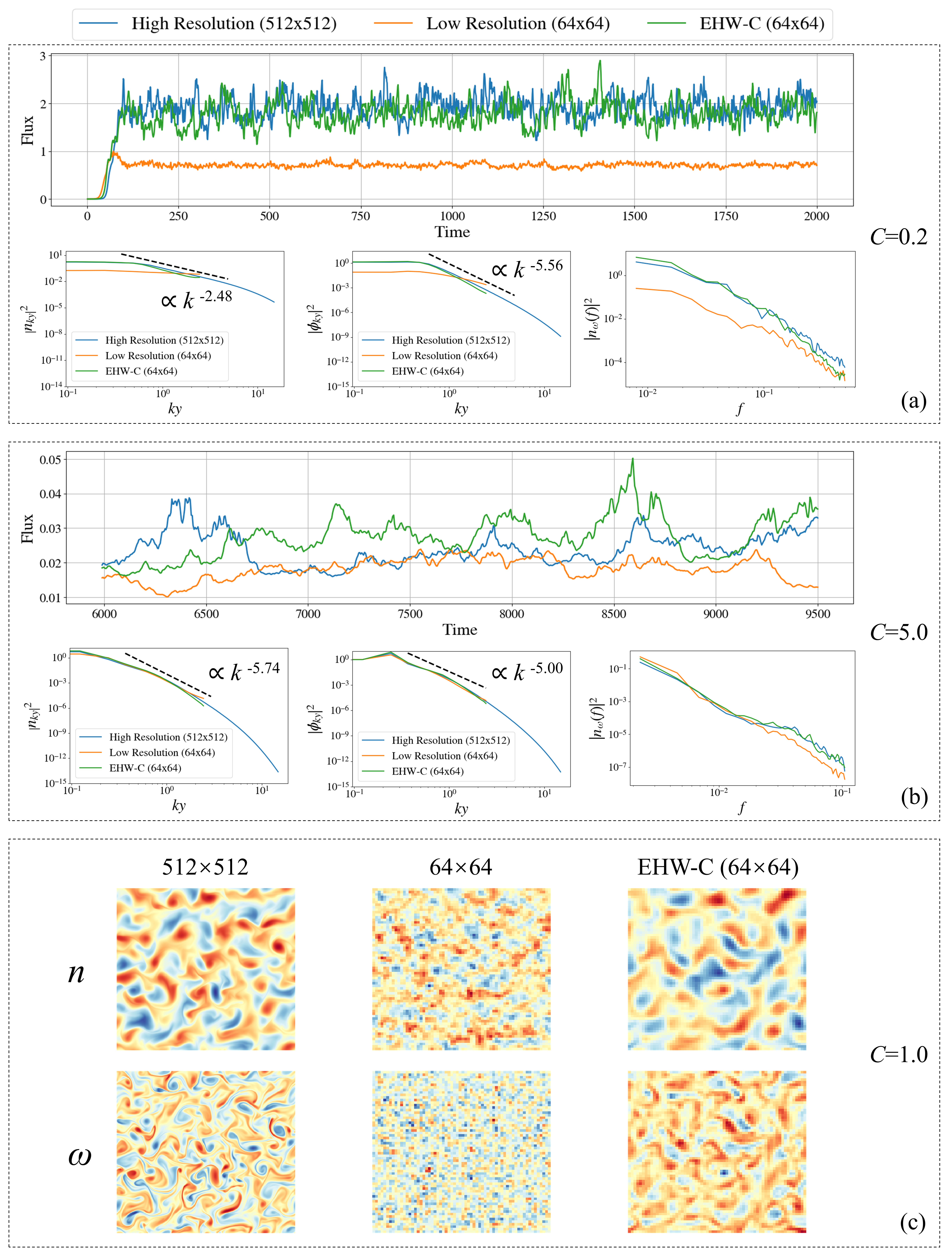}
\caption{Comparison of simulations under different resolutions. Subplot (a) and (b) compares the specific quantities at the limit cases, four key turbulent features are shown: the time evolution of particle flux, the density spectrum \(|n_{ky}|^2\), the potential spectrum \(|\phi_{k_y}|^2\), and the frequency spectrum \(|n_{\omega}(f)|^2\). Each plot compares the results from high resolution DNS (\(512\times512\)), low resolution DNS (\(64 \times 64\)), and the proposed EHW-C model at the same low resolution (\(64 \times 64\)). Subplot (c) compares the 2D plots of the case of $C=1$.}
\label{fig:flux_subfigs}
\end{figure}

In the near-adiabatic limit ($C=5.0$), the situation becomes more subtle.
Unlike the $C=0.2$ case, the spatial spectra show relatively similar behavior across all three cases, making it difficult to distinguish the closure effects through spectral analysis alone.
\added{Note that unlike the $C=0.2$ case, the contribution of $\phi_{k}$ becomes dominant, resulting in a steeper energy spectrum, which is because of inverse cascade of enstrophy.}
Additionally, the temporal flux evolution exhibits smaller differences among the three cases, compared to the $C=0.2$ case, with all simulations showing comparable flux saturation levels.
However, the frequency spectrum reveals crucial differences that are not apparent in the spatial domain.
As shown in Fig.~\ref{fig:flux_subfigs} (b), the frequency spectrum demonstrates that EHW-C model successfully captures the proper temporal dynamics of the turbulent fluctuations.
While the \deleted{unclosed} low resolution simulation fails to reproduce the correct frequency distribution, particularly at higher frequencies, EHW-C model closely follows the high resolution reference across the entire frequency range.
This indicates that although the spatial distribution may appear similar, the temporal characteristics of the turbulence are better captured by the closure model compared to the low-resolution case.
This frequency domain analysis highlights the importance of examining turbulent dynamics from multiple perspectives, as spatial spectra alone may not reveal the full impact of closure models, particularly in parameter regimes where the spatial energy cascade is less sensitive to resolution effects.

Across both regimes, EHW-C model achieves this level of accuracy on a significantly coarser spatial grid ($64 \times 64$) compared to the high resolution DNS benchmark ($512 \times 512$). 
This represents a 64-fold reduction in grid points, leading to over 90$\%$ reduction in computational time while preserving essential turbulent characteristics. 
The consistent performance across different physical regimes demonstrates the robustness and generalization capability of our physics-informed approach.

\subsection{Generalization Across Coefficient Space}\label{subsec2}

To explore the generalization capability of the identified closure formulation, we further investigate the performance of EHW-C model at intermediate adiabaticity parameter values that are not included in the original training data.
Specifically, we evaluate the model at \(C = \{0.5,1.5,2.5,3.5,4.0,4.5\} \) using the closure coefficients determined through interpolation from the trained values shown in Table~\ref{coefficients}.
The resulting coefficients are incorporated into EHW-C model and used in DNS on a low-resolution grid ($64\times64$).
The outputs of these simulations are then compared against high-resolution reference data ($512\times512$) in terms of physical quantities such as particle flux evolution, density spectra $|n_{k_{y}}|^{2}$, potential spectra $|\phi_{k_{y}}|^{2}$, and frequency spectra $|n_{\omega}(f)|^{2}$.

\begin{figure}[H]
\centering
\includegraphics[width=1\textwidth]{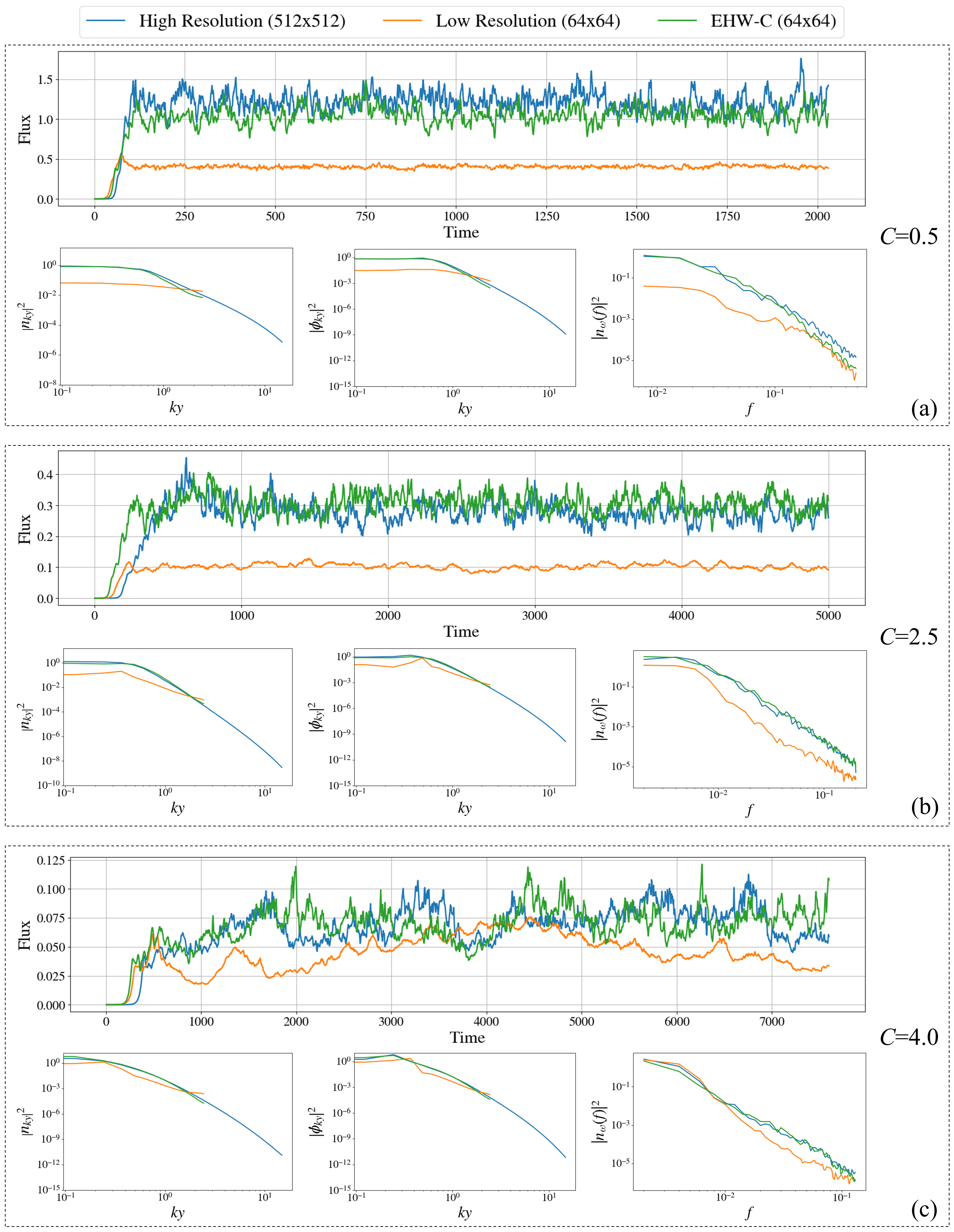}
\caption{Generalization performance of the proposed EHW-C model under unseen adiabaticity values \(C = \{0.5, 2.5, 4.0\} \). For each case, the six closure coefficients are interpolated from the coefficients in Table~\ref{coefficients}. The resulting simulations are conducted on a low resolution grid (\(64 \times 64\)) and compared with the high resolution reference results (\(512 \times 512\)) and the standard low resolution simulations without closure. Each row shows (from left to right): the time evolution of particle flux, the density spectrum \(|n_{k_y}|^2\), the potential spectrum \(|\phi_{k_y}|^2\) and frequency spectra $|n_\omega(f)|^2$.}
\label{test cases}
\end{figure}

Representative results are shown in Fig.~\ref{test cases}, with additional results provided in the supplementary material.
Across the intermediate values, EHW-C model consistently demonstrates \replaced{good agreement with high-resolution results}{excellent performance in reproducing the high-resolution benchmarks}.
The flux evolution shows that EHW-C model successfully captures the saturation behavior and temporal fluctuations that are severely underestimated by the \deleted{unclosed} low-resolution simulations.
In the spectral domain, EHW-C model accurately reconstructs the energy distribution across all wavenumbers for both density and potential spectra, maintaining the proper scaling relationships characteristic of each parameter regime.

Particularly noteworthy is the frequency spectrum analysis, which reveals that EHW-C model preserves the temporal dynamics across the entire frequency range, while \replaced{low resolution}{unclosed} simulations \deleted{completely} fail to capture these essential characteristics.
This consistent performance across different intermediate $C$ values demonstrates the robustness and strong generalization capability of the learned closure model, confirming that the interpolated coefficients maintain physical consistency across a continuous spectrum of adiabaticity values without requiring explicit training at every parameter point.

\section{Methods}\label{sec11}

\subsection{TOKAM2D}

TOKAM2D\cite{tokam2d2018,tokam2d2022,tokam2d_gyselax} is a GPU-based code which can solve Hasegawa-Wakatani equations\cite{bib3} in a 2D domain with periodic boundary conditions.
It employs a pseudo-spectral method based on the $2/3$ de-aliasing rule and uses a 4th-order Runge-Kutta scheme for time integration. 
Fast Fourier transforms serve as the primary computational operations for efficient spectral calculations.
The code can handle a wide range of adiabaticity parameters C, from the hydrodynamic limit ($C \ll 1$) to the adiabatic limit ($C\gg 1$).
The flexible architecture also allows for easy implementation of closure terms within the Hasegawa-Wakatani system.

\subsection{Direct Interaction Approximation}\label{subsec2}
Direct Interaction Approximation (DIA) is a statistical approach to describe the turbulence.
Time evolution of each mode is influenced by nonlinear interactions with other modes. 
When analyzing the time evolutions of these other modes by considering nonlinear interaction as well, this inevitably becomes an $n$-point problem.
DIA limits this to direct interactions up to the 3-point problem, treating each mode as follows.
First, it separates the Fourier component of turbulent fields as $\phi_{\mb{k}} = \bar{\phi}_{\mb{k}} + \tilde{\phi}_{\mb{k}}$, where $\bar{\phi}_{\mb{k}}$ is the mean part, which satisfies the maximal randomness condition, and $\tilde{\phi}_{\mb{k}}$ is the perturbed part, which contains the time memory of mean terms.
In this way, the 3 point correlation term can be written as follows:
\begin{eqnarray}
    \langle \phi_{\mb{k}}\phi_{\mb{p}}\phi_{\mb{q}}\rangle\simeq 
    \langle \bar{\phi}_{\mb{k}}\bar{\phi}_{\mb{p}}\tilde{\phi}_{\mb{q}}\rangle
    + \langle \bar{\phi}_{\mb{k}}\tilde{\phi}_{\mb{p}}\bar{\phi}_{\mb{q}}\rangle
    + \langle \tilde{\phi}_{\mb{k}}\bar{\phi}_{\mb{p}}\bar{\phi}_{\mb{q}}\rangle,
\end{eqnarray}
where $\langle \dots\rangle$ denotes a statistical average.
The notations for a direct interaction triad is $\mb{k} = \mb{p} + \mb{q}$.
The perturbed components are treated as non-Markovian processes to incorporate the temporal memory of mean field interactions, which has been discussed in Refs. \cite{Gang1990} and \cite{Gurcan2006}. 
Through this formulation, each three-point correlation term can be expressed as products of mean field components and response functions that capture the temporal memory effects. 
When applied to the Hasegawa-Wakatani system, this DIA procedure yields closure terms that appear as additional diffusion and hyperdiffusion operators, \added{as well as cross terms} in the Eqs. (\ref{hw_c1}) and (\ref{hw_c2}).
The closure coefficients $D_{nn}$, $D_{n\omega}$, $D_{\omega n}$, $D_{\omega\omega}$, $\mu_{n}$, and $\mu_{\omega}$ represent the collective influence of unresolved high wavenumber modes on the resolved scales.
These coefficients effectively capture the essential physics of turbulent energy transfer while maintaining computational tractability-a key advantage of the DIA approach for developing physically motivated closure models.

\subsection{Physics Informed Neural Networks Framework}\label{subsec2}

\begin{figure}[h]
\centering
\includegraphics[width=0.9\linewidth]{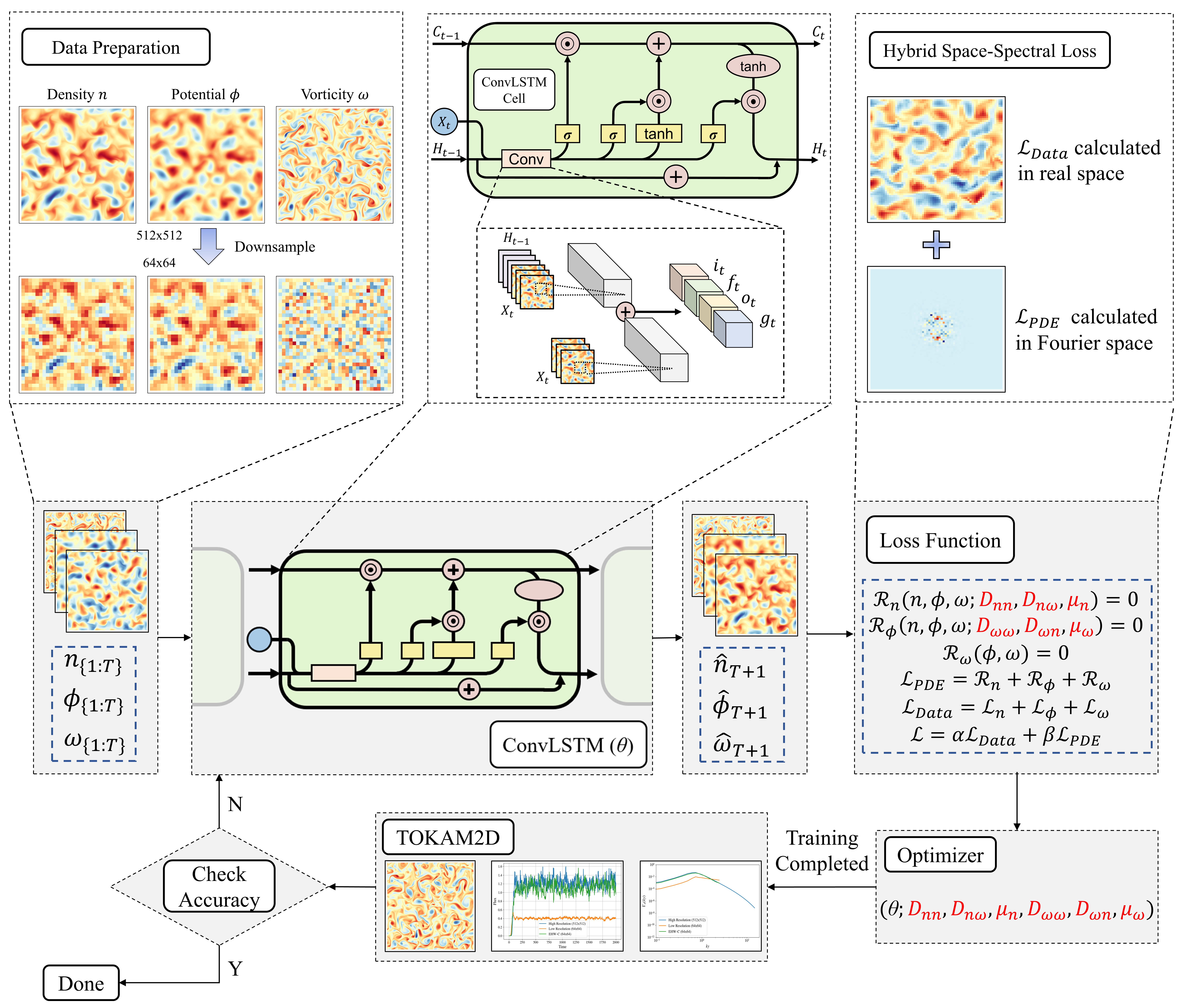}
\caption{Diagram of physics-informed networks framework for identifying closure coefficients in Hasegawa–Wakatani system. The workflow begins with high-resolution (512×512) simulation data of plasma density $n$, potential $\phi$, and vorticity $\omega$, which are down-sampled to 64×64 to generate low-resolution observations, which is used as the training data. These spatiotemporal fields are fed into a ConvLSTM network and the network is trained using a hybrid loss function composed of a data loss $\mathcal{L}_{\text{Data}}$ in real space and a physics-constrained PDE loss $\mathcal{L}_{\text{PDE}}$ computed in Fourier space. The PDE residuals $R_n$, $R_\phi$, and $R_\omega$ enforce consistency with the governing equations and explicitly depend on closure parameters $\{ D_{nn}, D_{n\omega}, \mu_n, D_{\omega\omega}, D_{\omega n}, \mu_\omega \}$. These coefficients are optimized jointly with the ConvLSTM weights to match both the data and the underlying physics. Model predictions are validated using full-resolution simulations from the reference solver TOKAM2D, ensuring that the learned closure captures the essential multiscale dynamics of plasma turbulence.}
\label{fig:pinn}
\end{figure}

The working schematic of the \added{Physics Informed Neural Networks} (PINNs) framework is illustrated in Fig.~\ref{fig:pinn}, which demonstrates how closure coefficients can be effectively identified using physics-informed learning. This approach can ensure that the identified closure coefficients are both physically consistent with the underlying governing laws and robustly applicable in coarse-grid turbulence modeling.

\subsubsection{Data Preparation}\label{subsubsec2} 

High-resolution training data is generated using TOKAM2D with a 512×512 grid over a domain of $L_x=L_y=51.5\rho_0$, where $\rho_0$ is the reference ion Larmor radius. 
The adiabatic coefficient $C$ is varied over the range $[0.2,5]$ to explore different turbulence regimes, which has been widely adopted in previous studies\cite{artigues2025accelerating,gahr2024scientific}.

To construct low-resolution training inputs, we adopt a down-sampling strategy that maps the high-resolution data to 64×64 grids by extracting every eighth point along each spatial dimension from the simulation fields (plasma density $n$, potential $\phi$, and vorticity $\omega$). This process creates the resolution mismatch that necessitates closure modeling, while preserving accurate values at the sampled locations.

The down-sampled data serves as input for training the PINNs framework to determine closure coefficients, enabling low-resolution simulations to approximate high-resolution behavior without requiring the full-resolution data during inference.

\subsubsection{Model Implementation Details}\label{subsubsec2} 
\deleted{The nature of the data—time series generated from numerical simulations—naturally motivates the use of a sequence modeling approach.} 
We benchmark three spatiotemporal \replaced{architecture}{backbones}—ConvLSTM, PredRNN, and TrajGRU—under a unified training protocol. To ensure fairness, all models share the same loss, data split, and optimization. For the specific training setup, all three candidate architectures—ConvLSTM, PredRNN, and TrajGRU—employ a stacked-cell design to enhance model capacity and enable hierarchical spatiotemporal feature extraction, consisting of four sequential cells with 64 hidden channels each and a kernel size of 3×3. The input at each time step is a three-channel two-dimensional field, representing the density $n$, potential $\phi$, and vorticity $\omega$. The model is trained using a sliding window approach, where each input sequence contains $T=30$ consecutive time steps and the target is to predict the state at time $T+1$. The full dataset comprises 1,500 time steps for each $C$ case, from which overlapping input-target pairs are extracted. The batch size is set to 30. Each input sample has a shape of [30,3,64,64], corresponding to a sequence of 30 frames of 3-channel spatial fields. The model was implemented using PyTorch 2.2.1 \cite{paszke2019pytorch}. All models were trained on an NVIDIA A100 GPU using the Adam optimizer \cite{kingma2014adam}. The training was conducted for a total of 1500 epochs. The initial learning rate was set to $1\times10^{-4}$, and a multi-step learning rate schedule was employed via a scheduler. Specifically, the learning rate was reduced by a factor of 0.5 at every 100 epochs, starting from epoch 100 until epoch 1500.

\subsubsection{Loss Function}\label{subsubsec2} 
To ensure that the neural network not only fits the data accurately but also adheres to the underlying physical laws, we design a hybrid space-spectral loss function that combines information from both real (physical) space and Fourier (spectral) space. Specifically, the total loss function consists of two components: a data loss and a physics-based PDE residual loss.

The training loss is expressed as a weighted combination of data loss $\mathcal{L}_{\text{data}}$ and PDE residual loss $\mathcal{L}_{\text{PDE}}$:
\begin{equation}
\mathcal{L} = \alpha \mathcal{L}_{\text{data}} + \beta \mathcal{L}_{\text{PDE}}
\label{eq:loss}
\end{equation}
where $\alpha$ and $\beta$ are tunable weights that balance the importance between data fidelity and physical consistency. The data loss $\mathcal{L}_{\text{data}}$ is computed in the real space by comparing the model predictions against the ground-truth simulation data. It enforces pointwise accuracy of the predicted density $n$, potential $\phi$, and vorticity $\omega$ fields. The PDE loss $\mathcal{L}_{\text{PDE}}$ is evaluated in Fourier space, where spatial differential operators (e.g., gradients, Laplacians) become simple algebraic multipliers. It is defined as the squared residuals of the governing equations, ensuring physical consistency by penalizing deviations from zero residuals in the frequency domain. This spectral formulation significantly simplifies the computation of high-order spatial derivatives, which are often numerically unstable or computationally expensive in real space. For example, second- or fourth-order derivatives required by closure terms can be efficiently and accurately implemented as multiplications by $k^2$ or $k^4$ in the Fourier domain, avoiding the need for repeated finite-difference or deep computational graphs introduced by automatic differentiation. The time derivatives in the equations are calculated by 4th order Runge-Kutta methods.

In addition, the advantage of introducing the PDE residual to the loss function is this design incorporates the six closure-related coefficients (i.e., $D_{nn}$, $D_{n\omega}$, $D_{\omega\omega}$, $D_{\omega n}$, $\mu_n$, and $\mu_\omega$) as explicit trainable variables within the loss function. Rather than treating these coefficients as fixed or externally prescribed quantities, we treat them as internal "weights" to be jointly optimized with the neural network weights. This design enables a learning process that not only fits the data and enforces PDE constraints, but also self-consistently identifies the optimal closure parameters from data.

\subsubsection{Verification and Validation of Identified Coefficients}\label{subsubsec2} 
To verify and validate the closure coefficients identified by the neural network, we do not rely on conventional quantitative metrics such as the coefficient of determination ($R^2$), root mean square error (RMSE), or other numerical evaluation criteria. Instead, we adopt a physics-based evaluation strategy where the identified coefficients are directly fed into the numerical solver (TOKAM2D). More specifically, we use DNS to solve EHW-C model with the inferred parameters on a low resolution grid, and the resulting simulation is taken as the sole criterion for assessment, to ensure that the learned coefficients are not only numerically reasonable but also physically meaningful.

The quality of the identified parameters is assessed based on their ability to reproduce key macroscopic and statistical features of the turbulent system. The simulation of the EHW-C model on a low resolution gird is compared high-resolution reference solutions. Only when the reproduced fields and spectra align well with the ground truth, the identified coefficients are considered valid and reliable. 

\subsection{Explore the Coefficient Space}\label{subsec2}
To investigate the generalizability of the identified closure model across different physical regimes, an exploration of the coefficient space with respect to the adiabatic coefficient $C$ is performed. Specifically, the obtained closure parameters (i.e., $D_{nn}$, $D_{n\omega}$, $D_{\omega\omega}$, $D_{\omega n}$, $\mu_n$, and $\mu_\omega$) for a representative set of $C$ values: \(C = \{0.2,1.0,2.0,3.0,5.0\} \) are used to fit a smooth polynomial function with respect to $C$, capturing their functional dependence on the adiabaticity.

Once these polynomial expressions are established, we use them to predict closure coefficients at intermediate values of $C$, specifically for \(C = \{0.5,1.5,2.5,3.5,4.0,4.5\} \). These predicted coefficients are then fed into the simulations via DNS on a low-resolution grid. To assess the validity of these predicted coefficients, the resulting low-resolution simulations are compared against high-resolution reference solutions in terms of key physical quantities such as density flux, energy spectra, and potential spectra. This procedure provides a systematic investigation for evaluating the interpolation capability and physical robustness of the learned closure model across a broader parameter space.

\section{Conclusion}\label{sec13}

Developing closure models that are simultaneously physically grounded and computationally tractable represents a fundamental challenge in plasma turbulence modeling. To address this challenge, we propose an Extended Hasegawa–Wakatani model with closure (EHW-C), which combines rigorous theoretical foundations with data-driven parameter identification. The construction of the EHW-C model follows a systematic two-stage methodology. First, we derive the explicit mathematical form of the closure terms based on \added{Direct Interaction Approximation (DIA) theory and eddy-damped quasinormal Markovian (EDQNM) approximation}. Second, we formulate the identification of closure coefficients as an inverse problem, which we solve using a \added{Physics Informed Neural Networks} (PINNs) framework that seamlessly integrates PDE residual loss with data-fitting loss terms. This unified approach enables direct determination of the optimal coefficient combination upon completion of model training. To ensure optimal network performance, we conducted a comprehensive comparative analysis of candidate backbone architectures, ultimately selecting ConvLSTM as the most suitable foundation based on its superior performance characteristics for this spatiotemporal modeling task.

During the verification and validation process, We directly verify the constructed model through high-fidelity simulations using the TOKAM2D solver. A key validation of our approach lies in the physical consistency of the inferred closure coefficients. Our PINNs framework successfully identifies negative closure coefficients in specific regimes—particularly negative vorticity diffusion ($D_{\omega\omega} < 0$) that introduces inverse cascade behavior characteristic of 2D fluid turbulence\cite{bib2}. Critically, when the system approaches the adiabatic limit ($C=5$), the EHW-C model reduces to the Hasegawa-Mima equation with closure terms that exhibit weak negative viscosity in the low-$k$ regime and positive hyperviscosity at high-$k$, in excellent agreement with the theoretical predictions of Smith and Hammett\cite{smith1997eddy}. This concordance between our data-driven coefficient identification and established theoretical expectations demonstrates that the PINNs framework has effectively learned the underlying physics rather than merely fitting data, providing confidence in the physical validity of the closure model across different turbulence regimes.

When validated in direct numerical simulations (DNS), EHW-C model accurately reproduces key dynamical signatures of high-resolution DNS—including particle flux and spectral features—while operating on a significantly coarser grid ($64 \times 64$ versus $512 \times 512$), resulting in over 90$\%$ reduction in computational cost. Particularly noteworthy is our frequency domain analysis, which reveals that EHW-C model captures temporal dynamics that remain invisible to spatial spectral analysis alone—a critical advantage for accurate transport predictions. Furthermore, by fitting the inferred coefficients as continuous functions of the adiabaticity parameter $C$, we enable model generalization across a broad range of turbulence regimes, including conditions unseen during training.

EHW-C model addresses key limitations in current closure modeling by bridging the gap between physical transparency and computational efficiency. Unlike purely data-driven approaches that lack interpretability, or conventional diffusive closures that miss essential turbulence physics, our framework combines theoretical rigor from DIA with data-driven flexibility through PINNs, maintaining both physical consistency and predictive accuracy.

Future research directions include extending this framework to three-dimensional geometries and more complex plasma models, such as gyrokinetic systems, to capture the full complexity of tokamak turbulence. Additionally, the integration of multi-scale physics and the development of closure models for edge plasma turbulence represent particularly promising avenues.

\backmatter

\bmhead{Supplementary information}

These figures provide additional results complementing Fig.~\ref{fig:flux_subfigs} and Fig.~\ref{test cases} in the main text, and is included here for completeness. It presents the time evolution of particle flux (top column) and the corresponding density (left column) and potential  (middle column) spectra for the other values of the adiabatic coefficient. Right column shows the frequency spectrum of the potential. The comparisons demonstrate that the proposed closure approach consistently recovers both flux levels and spectral shapes across different turbulent regimes.


\begin{figure}[h]
\centering
\includegraphics[width=1\linewidth]{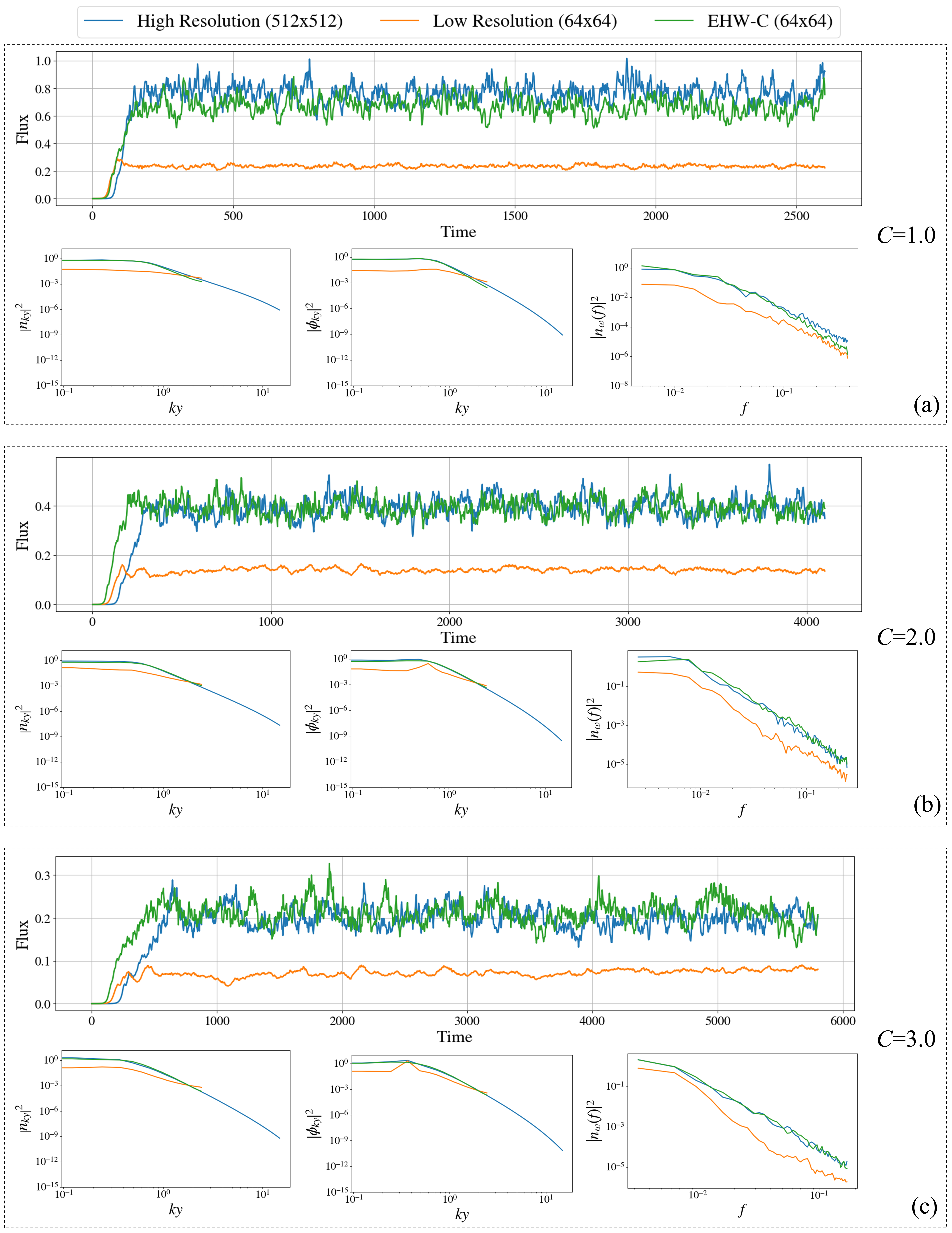}
\caption{Supplementary Figures for Fig.~\ref{fig:flux_subfigs}}
\label{fig:training case adx}
\end{figure}

\begin{figure}[h]
\centering
\includegraphics[width=1\linewidth]{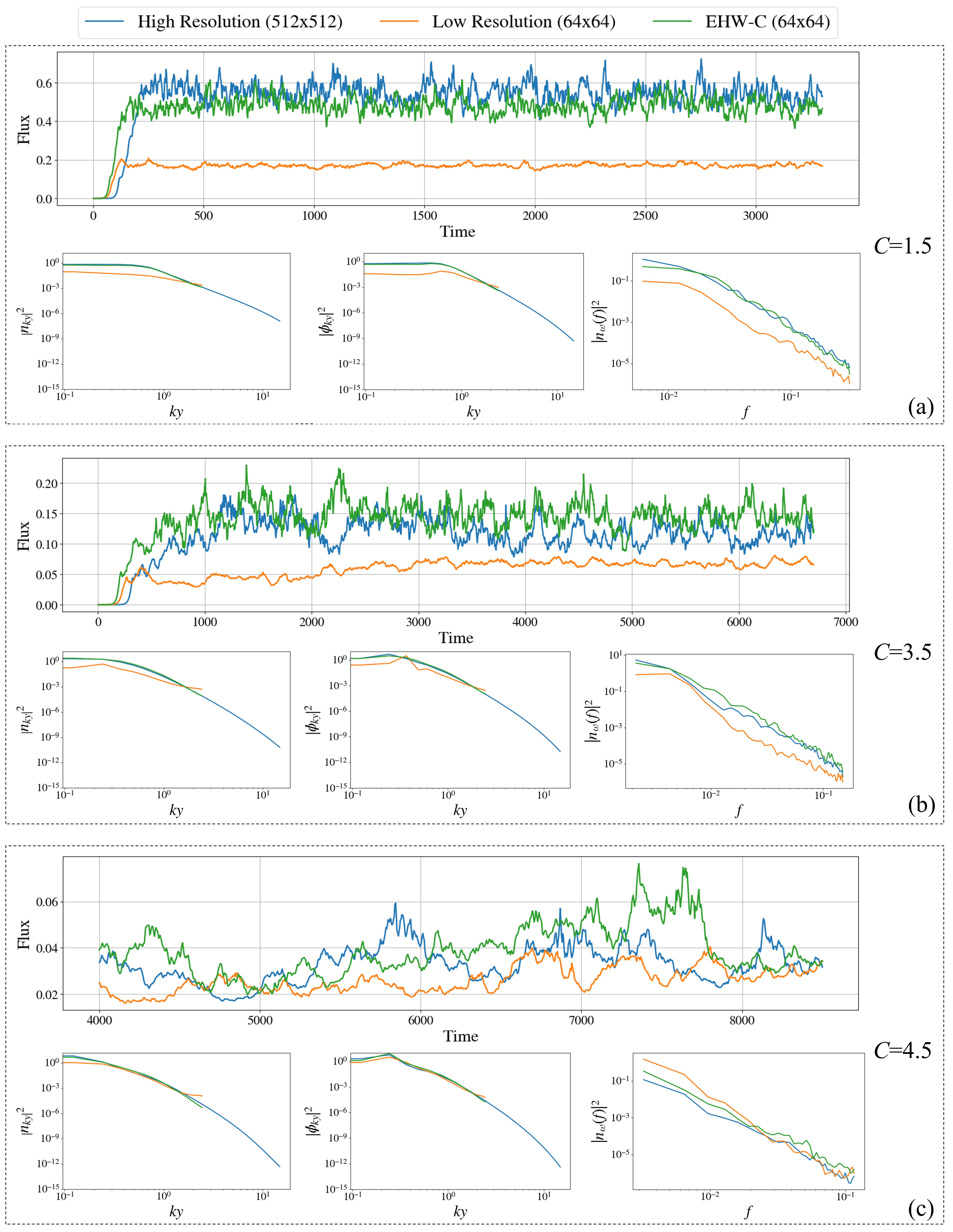}
\caption{Supplementary Figures for Fig.~\ref{test cases}}
\label{fig:test case adx}
\end{figure}

\bmhead{Acknowledgements}

The authors would like to thank to Dr. Liu Jiao, Dr. Chin Chun Ooi and Dr. Jian Cheng Wong for their technical assistance and insightful suggestions. The authors also acknowledge the support from the National Research Foundation, Singapore. The authors would like to acknowledge the SAFE team for providing access to the TOKAM2D code, which was essential for the numerical simulations carried out in this study; and would also like to acknowledge the organizers of the Festival de Théorie 2025 for providing a discussion environment that greatly benefited this work.

\section*{Declarations}

\begin{itemize}
\item Funding:
This research is supported by the National Research Foundation, Singapore.
\item Conflict of interest/Competing interests: 
The authors declare no conflict of interest
\item Data and Code availability: 
The data that supports the findings of this study and the corresponding code belong to the Singapore SAFE team/CEA and are available from the corresponding author upon reasonable request.
\end{itemize}

\bibliography{References}

\end{document}